\documentclass[twocolumn,showpacs,preprintnumbers,amsmath,amssymb]{revtex4}

\usepackage{graphicx}
\usepackage{dcolumn}
\usepackage{bm}

\begin{document}


\title{Evidence for infrared finite coupling in Sudakov resummation}

\author{Georges Grunberg}
\affiliation{%
Centre de Physique Th\'eorique de l'Ecole  
Polytechnique (CNRS UMR C7644),\\
        91128 Palaiseau Cedex, France
}%


\date{\today}

\begin{abstract}
New arguments are presented in favor of the infrared finite coupling approach to power
corrections in the context of Sudakov resummation. The more regular infrared behavior of some peculiar
combinations of Sudakov anomalous dimensions, free of Landau singularities at large $N_f$, is pointed out. A
general conflict between the infrared finite coupling and infrared renormalon approaches to power corrections is
explained, and a possible resolution is proposed, which makes use of the arbitrariness of the choice of
constant terms  in the Sudakov exponent. A simple ansatz for a `universal' non-perturbative Sudakov
effective coupling at large $N_f$ emerges  from these considerations. An alternative evidence for an
infrared finite {\em perturbative} effective coupling in the Drell-Yan process at large
$N_f$ (albeit at odds with the infrared renormalon argument) is found within the framework of Sudakov resummation
for eikonal cross sections of Laenen, Sterman and Vogelsang. 
\end{abstract}

\maketitle

\noindent The notion of an infrared (IR) finite coupling, and the related concept of universality,  to parametrize
power corrections in QCD has attracted much attention for a long time \cite{DMW}. In the present note I
display further evidence in favor of this assumption in the more specific framework of Sudakov resummation, taking
the examples of scaling violation in deep inelastic scattering (DIS) and in the Drell-Yan process (more details
can be found in \cite{Gru-talk}).

\noindent From the standard exponentiation formulas in Mellin $N$- space for the DIS structure function (I shall
adopt in general the notations of \cite{Vogt}), one gets  the  large $N$ relation

\begin{eqnarray}{d\ln F_2(Q^2,N)\over d\ln Q^2}=4 C_F\ H(Q^2)+\nonumber\\
4 C_F\int_{0}^1 dz{z^{N-1}-1 \over
1-z} A_{{\cal S}}[(1-z)Q^2]+{\cal O}(1/N)  
\label{eq:scale-viol},\end{eqnarray}
where $H(Q^2)$ is  given as a power series in
$a_s(Q^2)\equiv \alpha_s(Q^2)/4\pi$ with $N$-independent coefficients,
$4 C_F A_{{\cal S}}(k^2)= A(a_s(k^2))+dB(a_s(k^2))/d\ln
k^2$, and
$A$ (the universal  `cusp' anomalous dimension) and $B$ are  the standard Sudakov anomalous dimensions
relevant to DIS, given as  power series in
$a_s$.
$A_{{\cal S}}(k^2)$ shall be referred to as the `Sudakov effective coupling', a physical
`effective charge' \cite{GG} and we
 shall see (in the large $N_f$ limit) that it has drastically different IR properties then the individual
Sudakov anomalous dimensions it is composed of. 

\noindent The Borel transform $B[A_{{\cal S}}](u)$  of the Sudakov
effective coupling has been computed \cite{Gardi-Roberts} at large $N_f$. It is defined by
$A_{{\cal S}}(k^2)={1\over \beta_0}\int_0^\infty du
\exp(-u\ln k^2/\Lambda^2)  
 B[A_{{\cal S}}](u)$, where $\beta_0={11\over 3} C_2(G)-{2\over 3}N_f$ is the one-loop coefficient of the
beta function.  One gets, with the `na\"{\i}ve non-abelization' recipie: 
$B[A_{{\cal S}}](u)=\exp(-d u){\sin\pi u\over \pi u}\ {1\over
2}\left({1\over 1-u}+{1\over 1-u/2}\right)$, 
where  $d$ is a scheme-dependent constant related to the renormalization of fermion loops: $d=-5/3$
in the $\overline{MS}$ scheme and $d=0$ in the so called `V-scheme'. In the
following I shall use the `V-scheme' for simplicity. I note that $B[A_{{\cal S}}](u)$  is free of
renormalon singularities,
and  that for $k=\Lambda$ (the Landau pole of the one-loop V-scheme coupling) 
$A_{{\cal S}}(k^2=\Lambda^2)$ is finite. This behavior is in striking contrast 
with that of the cusp anomalous dimension \cite{Beneke-Braun} which displays  wild oscillations,
resulting in a completely unphysical behavior around
$k=\Lambda$. It is actually possible to get an analytic expression for
$A_{{\cal S}}(k^2)$, valid at all $k^2$, and one finds  that
$A_{{\cal S}}(k^2)$ approaches an {\em infinite} (and negative) IR fixed point for $k^2\rightarrow 0$: 
$A_{{\cal S}}(k^2)\simeq -{1\over 2\beta_0}{\Lambda^4\over k^4}+{1\over
2\beta_0}{\Lambda^2\over k^2}$.
This is not by itself  an unphysical behavior ( the coupling is causal and there is no Landau singularity),
except for  the  negative sign in the infrared, which cannot reproduce a vanishing Sudakov tail. The other 
trouble  is in the too strongly divergent IR behavior, which gives a  divergent  contribution  to the integral
on the right-hand side of eq.(\ref{eq:scale-viol}). These facts make  plausible the speculation
 that there exists a non-perturbative modification $\delta A_{{\cal S}}(k^2)$ of the coupling at 
small scales \cite{DMW} which might turn the infinite IR fixed point of perturbative origin into a genuinely
non-perturbative, but softer (eventually finite)  fixed point. 

\noindent However, there is a potential clash between the IR finite coupling  and the IR renormalons
approaches to power corrections (very closely connected to the well-known  issue \cite{GS},\cite{Beneke-Braun} 
of $1/Q$ corrections in Drell-Yan) which can be summarized  as follows. Consider a typical
`renormalon integral' 
$R(Q^2,N)=\int_{0}^{Q^2}{dk^2\over k^2} F(k^2/Q^2,N)
A_{{\cal S}}(k^2)$, 
where $F(k^2/Q^2,N)$ shall be referred to as the `Sudakov distribution function', and introduce its (`RS invariant'
\cite{Gru-Borel}) Borel representation    
$R(Q^2,N)={1\over \beta_0}\int_0^\infty du\exp(-u\ln Q^2/ \Lambda^2) B[R](u,N)$.
The Borel transform of $R$ is given by
$B[R](u,N)=B[A_{{\cal S}}](u)\times{\tilde F}(u,N)$,
with
${\tilde F}(u,N)=\int_0^1{dx\over x} F(x,N) \exp(-u\ln x)$.
The {\em factorized form} \cite{Gru-Borel},\cite{Beneke-Braun} of the expression should be noted. Suppose now
$B[A_{{\cal S}}](u)$ has a zero at $u=u_0$, which is 
{\em not shared} by
$B[R](u,N)$. Then  ${\tilde F}(u,N)$ must have a
pole 
at $u=u_0$. If
$u_0>0$, this means an IR renormalon, and $F(x,N)$ contains an ${\cal O }(x^{u_0})$ contribution for
$x\rightarrow 0$. If this renormalon is not present in $B[R](u,N)$, the standard IR renormalon philosophy
would conclude that no corresponding power correction is present. On the other hand, this power correction is
still expected in the IR finite coupling approach, where $A_{{\cal S}}(k^2)$ is assumed to have a finite IR
fixed point and the low energy part of the renormalon integral $R(Q^2,N)$ is well-defined. Then the
distribution function
$F(k^2/Q^2,N)$ can be expanded in powers of $k^2$ for $k^2\rightarrow 0$, yielding a non-vanishing power
correction for each term \cite{footnote2} in its low energy expansion, parametrized by a low energy moment of the
effective coupling
$A_{{\cal S}}(k^2)$. 
Even if
$B[R](u,N)$ is singular, rather than non-vanishing and finite, at
$u=u_0$, there is still a clash, since the singularity of ${\tilde F}(u,N)$ is necessarily stronger than that of
$B[R](u,N)$ in presence of a zero of $B[A_{{\cal S}}](u)$; thus the IR finite coupling approach will predict a 
{\em more enhanced} power correction than indicated by the renormalon argument.
Two examples of this situation at large $N_f$ are:

\noindent i) {\em DIS}: there   $B[A_{{\cal S}}](u)$ vanishes for any integer
$u$,
$u\geq 3$.

\noindent ii) {\em Drell-Yan}: in this case the Sudakov effective coupling which occurs in the $Q^2$ derivative of
the Drell-Yan cross section is given by \cite{footnote3}
$4 C_F A_{{\cal S},DY}(k^2)=  A(a_s(k^2))+{1\over 2}dD_{DY}(a_s(k^2))/ d\ln
k^2$,
and the corresponding Borel transform at $N_f=\infty$ is \cite{Beneke-Braun} (dropping an inessential
 $\exp(c u)$ factor which corresponds  to a change of scale):
$B[A_{{\cal S},DY}](u)= {1\over \Gamma(1+u)} {\pi^{1/2}\over \Gamma(1/2-u)}$,
which vanishes for positive half-integer $u$.
In both cases, the zeroes are absent from the corresponding Sudakov exponent, and the IR finite coupling
approach will lead to the prediction of extra power corrections in the exponent ($N^3/Q^6$, $N^4/Q^8$, etc...in
DIS, and $N/Q$, $N^3/Q^3$, etc...in Drell-Yan).
To make progress, we have to understand better the origin of zeroes in  
$B[A_{{\cal S}}](u)$. 

\noindent\underline {DIS case}: putting
$S(Q^2,N)=\int_{0}^1 dz{z^{N-1}-1 \over 1-z} A_{{\cal S}}[(1-z)Q^2]$,
it is useful to write $S(Q^2,N)$ 
as a `renormalon integral', where
the Sudakov distribution function is  given by
$F(k^2/Q^2,N)=(1-k^2/Q^2)^{N-1}-1.$
Order by order the perturbative series of $S(Q^2,N)$ contain both ${\cal O}(N^0)$ `constant
terms' and terms which vanish as $N\rightarrow\infty$.
A meaningful simplification is achieved  by making use
of the following important scaling property:  
defining $\epsilon=N k^2/ Q^2$, and
$F(k^2/Q^2,N)\equiv G(\epsilon,N)=(1-\epsilon/N)^{N-1}-1$,
and taking the limit $N\rightarrow \infty$  with {\em $\epsilon$  fixed} one gets a
{\em finite} result
$G(\epsilon,N)\rightarrow G(\epsilon,\infty)\equiv G(\epsilon)$
with
$G(\epsilon)=\exp(-\epsilon)-1$.
 Let us now redefine the Sudakov exponent by using $
G(\epsilon)$ as the new distribution function, i.e. $S(Q^2,N)$ is replaced by
$S_{stan}(Q^2,N)=\int_{0}^{Q^2}{dk^2\over k^2} G(N k^2/Q^2)
A_{{\cal S}}(k^2)$ where the subscript `stan' means `standard'. This step is legitimate since, order by order in
perturbation theory, $S_{stan}(Q^2,N)$ and
$S(Q^2,N)$ differ only \cite{Gru-talk} by terms which vanish as $N\rightarrow\infty$, and thus share the {\em
same} 
$\ln N$ and constant terms.  I next assume a similar  ansatz for a modified Sudakov exponent
\begin{equation}S_{new}(Q^2,N)=\int_{0}^{Q^2}{dk^2\over k^2} G_{new}(N k^2/Q^2)
A_{{\cal S}}^{new}(k^2)\label{eq:ren-int-new},\end{equation}
and show that a {\em unique} solution for $G_{new}(N k^2/Q^2)$ and $A_{{\cal S}}^{new}(k^2)$ exists, under the
condition that $S_{new}$ reproduces all divergent $\ln N$ terms, together with an (a priori {\em arbitrary})  given
set of constant terms. The latter are usually  a subset of the set of all constant terms on the right hand side of 
eq.(\ref{eq:scale-viol}), which are uniquely defined, but (arbitrarily)  split between those included in the
Sudakov exponent $S(Q^2,N)$, and those belonging to $H(Q^2)$.  This statement can be checked order by order in
perturbation theory. Here I give an all-order proof. I first introduce
 a simplification appropriate to the large N limit. The perturbative series
  of $S_{new}(Q^2,N)$  still contains ${\cal O}(1/N)$
terms at large $N$. It is useful to discard them, and find a Borel representation of the corresponding series 
$S_{new}^{as}(Q^2,N)$ 
  which contains only the (same)
$\ln^p N$ and ${\cal O}(N^0)$ terms, with all the ${\cal O}(1/N)$ terms dropped.
One can show the looked for Borel transform is given by 

\begin{eqnarray}B[S_{new}^{as}](u,N)=B[A_{{\cal S}}^{new}](u) \nonumber\\
\times\left[ \exp(u\ln N)
\int_0^{\infty} {d\epsilon\over \epsilon}  G_{new}(\epsilon) \exp(-u\ln
\epsilon)- {G_{new}(\infty)\over  u}\right]\label{eq:B-as-disp-ren}\end{eqnarray}
where  $G_{new}(\infty)=-1$ is a subtraction term, and is fixed, since it determines the
leading logs. Eq.(\ref{eq:B-as-disp-ren}) displays the general form of the large
$N$ Borel transform:

\begin{equation}B[S_{new}^{as}](u,N)=\exp(u\ln N) B(u)+{C_{new}(u)\over u}\label{eq:B-as-form}.\end{equation}
It is clear that the $N$-dependent part $B(u)$ of the  Borel transform is unambiguously determined, and
cannot be changed without changing the $N$-dependent terms, i.e. the coefficients of all logs of
$N$, which are fixed. On the other hand, the  constant terms included into the Sudakov exponent are
obtained by setting
$N=1$ in eq.(\ref{eq:B-as-disp-ren}) or (\ref{eq:B-as-form}), and can be changed only by modifying the function
$C_{new}(u)$ (with
$C_{new}(0)=1$). Given  $B(u)$ from an independent calculation, and given an (a priori {\em arbitrary}) function
$C_{new}(u)$, eq.(\ref{eq:B-as-disp-ren}) and (\ref{eq:B-as-form}) determine in principle {\em both} $B[A_{{\cal
S}}^{new}](u)=C_{new}(u)$ and $G_{new}(\epsilon)$, which completes the  proof.

\noindent\underline {Large $N_f$ result}: Specializing now to large $N_f$ to compute $B(u)$ (and eventually
$C_{new}(u)$) , the result for the Borel transform of
$d\ln F_2(Q^2,N)/d\ln Q^2\vert _{SDG}$ at {\em finite}
$N$ can be given
 in the  `massive',  `single dressed gluon' (SDG)   dispersive Minkowskian formalism
\cite{BB},\cite{DMW} as 

\begin{eqnarray}B[d\ln F_2(Q^2,N)/d\ln Q^2]_{SDG}(u,N)=&4 C_F {\sin\pi u\over\pi u}\nonumber\\
\times\int_0^{\infty} {dx\over x} \ddot{{\cal F}}_{SDG}(x,N) \exp(-u\ln x) &\label{eq:B-SDG},\end{eqnarray}
where the `characteristic function' ${\cal F}_{SDG}(x,N)$ has been computed in \cite{DMW} 
($x=\lambda^2/Q^2$ where $\lambda$ is the `gluon mass'). I have checked that here too a similar scaling property
holds at large
$N$, namely, putting
${\cal G}_{SDG}(y,N)\equiv{\cal F}_{SDG}(x,N)$ with $y\equiv N x=N \lambda^2/Q^2$, one gets for
$N\rightarrow\infty$ at fixed $y$:
$\ddot{{\cal G}}_{SDG}(y,N)\rightarrow \ddot{{\cal G}}_{SDG}(y,\infty)\equiv
\ddot{{\cal G}}_{SDG}(y)$,
with

\begin{eqnarray} \ddot{{\cal G}}_{SDG}(y)=-1+ \exp(-y)-{1\over 2} y\ \exp(-y)\nonumber\\
-{1\over 2} y\
\Gamma(0,y)+{1\over 2} y^2\
\Gamma(0,y)
\label{eq:Gdot-SDG},\end{eqnarray}
where $\Gamma(0,y)$ is the incomplete gamma function and implies $\int_0^{\infty} {dy\over y} \ddot{{\cal
G}}_{SDG}(y)
\exp(-u\ln y)=\Gamma(-u){1\over 2}\left({1\over 1-u}+{1\over 1-u/2}\right)$. Thus for $N\rightarrow\infty$

\begin{eqnarray}B[d\ln F_2(Q^2,N)/d\ln Q^2]_{SDG}^{as}(u,N)=4 C_F {\sin\pi u\over\pi
u}\nonumber\\
\times\left[
\exp(u\ln N)
\int_0^{\infty} {dy\over y} \ddot{{\cal G}}_{SDG}(y) \exp(-u\ln y)
+{\Gamma_{SDG}(u)\over u}\right]\nonumber\\
\label{eq:B-SDG-as-disp-ren}\end{eqnarray}
where $\Gamma_{SDG}(u)=\left({\pi u\over\sin\pi
u}\right)^2{1\over (1-u)(1-u/2)}$ is again a subtraction function which takes into account {\em all}
constant terms of the SDG result.   The latter are indeed obtained by setting
$N=1$ in eq.(\ref{eq:B-SDG-as-disp-ren}). Thus, if one wants to select as input an {\em arbitrary} subset of
constant terms to be included into a {\em new} asymptotic Sudakov exponent $S_{new}^{as}(Q^2,N)$, one should
simply change the subtraction function
$\Gamma_{SDG}(u)$, namely define a new input $B[d\ln F_2(Q^2,N)/d\ln
Q^2]_{SDG}^{as,new}$ by eq.(\ref {eq:B-SDG-as-disp-ren}), with $\Gamma_{SDG}(u)\rightarrow\Gamma_{new}(u)$
where $\Gamma_{new}(u)$ (with $\Gamma_{new}(0)=1$) takes into account the new  set of
constant terms.  Having no ${\cal
O}(1/N)$ terms, one can now identify $4 C_F B[S_{new}^{as}](u,N)$ with $B[d\ln F_2(Q^2,N)/d\ln
Q^2]_{SDG}^{as,new}(u,N)$,
which yields the {\em two} master equations

\begin{eqnarray} B[A_{{\cal S}}^{new}](u) \int_0^{\infty} {d\epsilon\over \epsilon}
  G_{new}(\epsilon) \exp(-u\ln \epsilon)\equiv B(u)\nonumber\\
={\sin\pi u\over\pi u}\int_0^{\infty}{dy\over y} \ddot{{\cal G}}_{SDG}(y)
\exp(-u\ln y)\label{eq:A-G-SDG},\end{eqnarray}
and
\begin{equation} B[A_{{\cal S}}^{new}](u)\equiv C_{new}(u)=
\left({\sin\pi u\over\pi u}\right) \Gamma_{new}(u)\label{eq:A-G}.\end{equation}
Eq.(\ref{eq:A-G-SDG}) and (\ref{eq:A-G}) allow to determine {\em both} the Sudakov distribution
function
$ G_{new}(\epsilon)$ and the Borel transform of the associated Sudakov effective coupling $B[A_{{\cal
S}}^{new}](u)$, for a given input set of  constant terms which fix the subtraction
function  $\Gamma_{new}(u)$. It is interesting that $B[A_{{\cal S}}^{new}](u)$ is given {\em entirely}
by the subtraction term. 
Applying these results to the standard exponentiation formula, where
$G(\epsilon)$ is known, and using
eq.(\ref{eq:A-G-SDG}), one can in particular rederive  the previously quoted  result for $B[A_{{\cal S}}](u)$. I
note that the simplest possible ansatz, namely $\Gamma_{simple}(u)\equiv 1$,  gives
$B[A_{{\cal S}}^{simple}](u)={\sin\pi u\over\pi u}$, hence 
\begin{equation}A_{{\cal
S}}^{simple}(k^2)={1\over\beta_0}\left[{1\over
2}-{1\over\pi}\arctan(t/\pi)\right]\label{eq:A-simple},\end{equation}
with $t=\ln k^2/\Lambda^2$. It turns out that
$A_{{\cal S}}^{simple}(k^2)$ is already causal \cite{Gru-talk} and  IR finite (with $A_{{\cal
S}}^{simple}(0)=1/\beta_0$) at the {\em perturbative} level, and no non-perturbative modification is a priori
necessary.
However, the first two zeroes at
$u=1$ and
$u=2$  lead to two leading {\em log-enhanced} ${\cal O}({N\over Q^2}\ln Q^2)$ and ${\cal O}({N^2\over Q^4}\ln Q^2)$
power corrections at large $N$ in the IR finite coupling framework, as can be easily checked since we have in
this case
$G_{simple}(\epsilon)=\ddot{{\cal G}}_{SDG}(\epsilon)
\simeq {1\over 2}[\epsilon(\ln\epsilon+\gamma_E-3)+\epsilon^2(-\ln\epsilon-\gamma_E+1)+{\cal O}(\epsilon^3)]$. 
There is thus, as expected, a discrepancy with the IR renormalon expectation, which predicts only
 ${\cal O}({N\over Q^2})$ and ${\cal O}({N^2\over Q^4})$
power corrections.

\noindent\underline{Drell-Yan case}: 
The analogue of eq.(\ref{eq:scale-viol}) for the scaling violation of the short distance Drell-Yan cross-section
is $d\ln \sigma_{DY}(Q^2,N)/ d\ln Q^2=4 C_F\left[ H_{DY}(Q^2)+S_{DY}(Q^2,N)\right]+{\cal O}(1/N)$
with $S_{DY}(Q^2,N)=\int_{0}^1 dz\  2{z^{N-1}-1 \over
1-z} A_{{\cal S},DY}[(1-z)^2 Q^2]$.  
$S_{DY}(Q^2,N)$ can again be written as a `renormalon integral', with
$F(k^2/Q^2,N)\rightarrow F_{DY}(k^2/Q^2,N)=(1-k/Q)^{N-1}-1$ and $A_{{\cal S}}(k^2)\rightarrow A_{{\cal
S},DY}(k^2)$. The Sudakov distribution function
$F_{DY}(k^2/Q^2,N)$
involves {\em both} \cite{GS} even and odd powers of $k$ at small $k$. Taking the scaling limit
$N\rightarrow\infty$ with $\epsilon_{DY}=Nk/Q$ fixed one thus gets 
$G_{DY}(\epsilon_{DY})=\exp(-\epsilon_{DY})-1$.
Now the work of \cite{Vogelsang} for eikonal cross sections suggests to  use instead (thus avoiding odd power
terms)
$S_{DY}^{new}(Q^2,N)=\int_{0}^{Q^2}{dk^2\over k^2} G_{DY}^{new}(N k/Q)
A_{{\cal S},DY}^{new}(k^2)$,
with a  distribution function (I deal with the log-derivative of the Drell-Yan cros-section)
\begin{equation}G_{DY}^{new}(N k/Q)=2 {d\over d\ln Q^2}\left[K_0(2
N k/Q)+\ln(N k/Q)+\gamma_E\right]\label{eq:G-DY-new},\end{equation}
i.e.
 $G_{DY}^{new}(\epsilon_{DY})=-\left[1+x{dK_0\over dx}(x=2
\epsilon_{DY})\right]$,
where $K_0(x)$ is the modified Bessel function of the second kind. I note that
$G_{DY}^{new}(\epsilon_{DY})\rightarrow -1$ for $\epsilon_{DY}\rightarrow \infty$, consistently with the large
$\epsilon_{DY}$ limit of $G_{DY}(\epsilon_{DY})$. 
At large $N_f$  one gets, instead  of eq.(\ref{eq:A-G-SDG}) (eq.(\ref{eq:A-G}) remains the same)

\begin{eqnarray}B[A_{{\cal S},DY}^{new}](u) \int_0^{\infty}
2{d\epsilon\over\epsilon} G_{DY}^{new}(\epsilon) \exp(-2 u\ln\epsilon)=\nonumber\\
{\sin\pi u\over\pi
u}\int_0^{\infty}{dy\over y} \ddot{{\cal G}}_{SDG,DY}(y)
\exp(-u\ln y)\label{eq:A-G-SDG-DY},\end{eqnarray}
where $y=N^2 \lambda^2/Q^2$.
On the other hand, the large $N_f$ calculation of \cite{Beneke-Braun} yields the $N$-dependent part of the
large $N$ Borel transform
\begin{equation}{\sin\pi u\over\pi
u}\int_0^{\infty}{dy\over y} \ddot{{\cal G}}_{SDG,DY}(y)
\exp(-u\ln y)=-{1\over u}{\Gamma(1-u)\over \Gamma(1+u)}\label{eq:B-SDG-DY},\end{equation}
whereas eq.(\ref{eq:G-DY-new}) gives 
\begin{equation}\int_0^{\infty}
2{d\epsilon\over\epsilon} G_{DY}^{new}(\epsilon) \exp(-2 u\ln\epsilon)=-u
[\Gamma(-u)]^2\label{eq:B-K0}.\end{equation} 
From eq.(\ref{eq:A-G-SDG-DY}) one thus derives the large $N_f$ result
\begin{equation}B[A_{{\cal S},DY}^{new}](u)={1\over
\Gamma(1+u)\Gamma(1-u)}\equiv{\sin\pi u\over\pi u}\label{eq:A-new-DY},\end{equation}
hence 
$A_{{\cal S},DY}^{new}(k^2)= A_{{\cal S}}^{simple}(k^2)$. Thus in the framework of \cite{Vogelsang} the
{\em simplest} IR finite {\em perturbative} coupling naturally emerges! The fact that the new Sudakov distribution
function 
$ G_{DY}^{new}(\epsilon)$ implies a new Sudakov effective
coupling is of course one of the main point of the present paper. Again, there is a discrepancy \cite{footnote}
 with the IR
renormalon expectation: although   $G_{DY}^{new}(\epsilon_{DY})\simeq \epsilon_{DY}^2(2\ln\epsilon_{DY}+
2\gamma_E-1)+{\cal O}(\epsilon_{DY}^4\ln\epsilon_{DY})$ 
involves only {\em even} powers of $\epsilon_{DY}$ for
$\epsilon_{DY}\rightarrow 0$, they are logarithmically enhanced. One thus gets  ${\cal O}({N^{2p}\over
Q^{2p}}\ln Q^2)$ power corrections ($p$ integer) at large $N$, instead of the ${\cal O}({N^{2p}\over Q^{2p}})$
corrections expected from IR renormalons.

\noindent\underline{Large $N_f$ ansatz for a `universal' Sudakov  coupling}:
the question  arises whether it is possible to reconcile
the IR renormalon and IR
finite coupling approaches.  Actually, I should first stress it is not yet clear whether the
two approaches should be necessarily reconciled. For instance, in the DIS case, it could be that the operator
product expansion (OPE) at large
$N$ is consistent  with the existence of higher order power
corrections ($N^3/Q^6$, $N^4/Q^8$,...)  {\em in the exponent}. If this turns out to
be the case,  the IR finite coupling approach would be consistent
with the OPE (and at odds with the IR renormalon prediction) even  within the standard \cite{Vogt} exponentiation
framework. Similarly, the ansatz
$A_{{\cal S}}^{simple}(k^2)$ might be the correct one in the Drell-Yan case \cite{Vogelsang},  although it also
contradicts the  IR renormalon expectation. It is interesting to note   that eq.(\ref{eq:A-G}) indicates that
zeroes in
$B[A_{{\cal S}}^{new}](u)$ arise from {\em two} distinct sources: either the `universal' $\sin\pi u/\pi u$ factor
({\em simple} zeroes at integer
$u$ can come only from there), or the `arbitrary' $ \Gamma_{new}(u)$ subtraction term (zeroes at
half-integer $u$ can come only from there).
The previous discussion  suggests that  zeroes coming from the `universal'
$\sin\pi u/\pi u$ factor need {\em not} be necessarily removed in the IR finite coupling approach, at the
difference of the more `artificial' zeroes (such as those occuring in the standard Drell-Yan case) coming from the
subtraction term.

\noindent Notwithstanding the above remarks, I shall adopt in the following the attitude that the IR renormalon and
IR finite coupling approaches to power corrections should {\em always} be made consistent with one another.
For this purpose, one  must 
remove {\em all}
 zeroes from
$B[A_{{\cal S}}^{new}](u)$. The {\em mathematically simplest} `universal'  ansatz suggested
by eq.(\ref{eq:A-G}) is to choose
$\Gamma_{new}(u)=\Gamma(1-u)$,
which yields
$B[A_{{\cal S}}^{new}](u)=\left({\sin\pi u\over\pi u}\right)
\Gamma(1-u)={1\over \Gamma(1+u)}$.
It is interesting to compare this result with the one obtained by eliminating the half-integer zeroes from the
standard Drell-Yan  result. There the simplest procedure is to  define
$B[A_{{\cal S},DY}^{new}](u)= {\Gamma(1/2-u)\over
\pi^{1/2}}B[A_{{\cal S},DY}](u)$,
which yields again the previous  ansatz.
Given  $B[A_{{\cal S}}^{new}](u)$, eq.(\ref{eq:A-G-SDG}) or (\ref{eq:A-G-SDG-DY})  determine  the Sudakov
distribution function
$G_{new}(\epsilon)$. In particular, assuming the above
 ansatz, one gets in the DIS case 

\begin{eqnarray}G_{new}(\epsilon)&=&-{\epsilon(1+\epsilon)\over 2}\ \  0\leq\epsilon\leq 1\nonumber\\
&=&-1\ \ \ \ \ \ \ \ \ \ \ \  \epsilon\geq 1\label{eq:Gnew},\end{eqnarray}
where $\epsilon=N k^2/Q^2$. Thus all power corrections beyond the two leading ones are indeed absent from the
Sudakov exponent, in agreement with the renormalon argument. In the Drell-Yan case, one gets instead

\begin{equation}G_{DY}^{new}(\epsilon_{DY})=\exp(-\epsilon_{DY}^2)-1\label{eq:Gnew-DY},\end{equation}  
i.e. only even powers of $\epsilon_{DY}=Nk/Q$, with no logarithmic enhancement for $\epsilon_{DY}\rightarrow 0$.
There is compelling numerical evidence that the  corresponding `universal'
 Sudakov effective coupling:
$A_{{\cal S}}^{new}(k^2)={1\over\beta_0}\int_0^\infty du
\exp\left(-u\ln k^2 /\Lambda^2\right)\
1/\Gamma(1+u)$
 blows up very fast (but remains positive) for $k^2\rightarrow 0$:
$A_{{\cal S}}^{new}(k^2)\simeq {1\over \beta_0}\exp\left({\Lambda^2\over k^2}\right)$.
Assuming that this behavior is indeed correct, one can again speculate  that there exists a
non-perturbative modification $\delta A_{{\cal S}}^{new}(k^2)$ of the coupling at small scales which will
generate a  non-perturbative but 
finite IR fixed point (a simple ansatz for $\delta
A_{{\cal S}}^{new}(k^2)$ has been given in \cite{Gru-talk}).

\noindent To conclude, I have shown that in a number of examples at large $N_f$ the Sudakov effective couplings
display remarkably smooth IR properties, with no Landau singularities (a rather unusual occurence in resummed
perturbation theory). Generalizing from the present findings, one can  distinguish two main possibilities. Either
the subtraction term
$\Gamma_{new}(u)$ has some singularities at  finite and positive values of $u$ (which must be simple poles
for the corresponding Sudakov coupling to be free of IR renormalons): this is the case in DIS and Drell-Yan with
the standard choice of the Sudakov distribution function;  or it has  no $u>0$ singularities, as  the `simple'
coupling eq.(\ref{eq:A-simple}), which emerges naturally in  the formalism of
\cite{Vogelsang} for the Drell-Yan cross-section. In the former case, the corresponding Sudakov effective coupling
is presumably \cite{footnote4} causal, but strongly IR divergent, requiring a low scale non-perturbative
modification to build  an IR finite coupling. In the latter case, the Sudakov coupling is likely (provided
$\Gamma_{new}(u)$ does not decrease faster then an exponential for $u\rightarrow\infty$) to be already  IR finite
(and causal) at the {\em perturbative} level. Even then, there may be extra non-perturbative effects,  which could
also take the form of a non-perturbative modification of the  coupling at low scales. I consider these facts as
hints strongly supporting the IR finite coupling hypothesis. The above ansatzes are in conflict with the IR
renormalon expectations for power corrections. The simplest attempt to reconcile the two approaches leads to the
proposal of a {\em universal} perturbative Sudakov coupling, which again requires a non-perturbative modification
to achieve IR finitness. The latter could introduce some amount of non-universality. The variety of Sudakov
distribution functions showing up in the exponentiation procedure is related to the freedom to select an arbitrary
set of `constant terms'  in the Sudakov exponent. This freedom  leaves open the question of the
determination of the `correct' Sudakov distribution function. There is no obvious way to solve this ambiguity. In
particular, one can show
\cite{Gru-talk} that the natural option of including {\em all} ${\cal O}(N^0)$ terms into the Sudakov exponent is
excluded in the DIS case: at large
$N_f$, there is no solution for the corresponding Sudakov distribution function, if one defines the Sudakov
effective coupling by eq.(\ref{eq:A-G}) with $\Gamma_{new}(u)\rightarrow\Gamma_{SDG}(u)$ (moreover this
effective coupling turns out to have renormalons). In the Drell-Yan case, the corresponding Sudakov distribution
function does exist, but implies ${\cal O}(1/Q)$ power corrections (and renormalons in the associated effective
coupling). With the option to incorporate {\em all} constant terms into the Sudakov exponent disfavored, there
remains the alternative to absorb into the  exponent a `minimum' set of constant terms, realized
\cite{footnote5} by the (universal) choice
$\Gamma_{new}(u)=1$  corresponding to the  `simple' coupling eq.(\ref{eq:A-simple}). A 
variant of this `minimal' option, inspired by the work of \cite{Magnea}, would allow to correlate the choice of
`left-over' constant terms functions $H_{new}$ and $H_{DY}^{new}$ not included in the  exponents $S_{new}$
and
$S_{DY}^{new}$   to the absolute ratio
$\vert{{\cal F}(-Q^2)\over{\cal
F}(Q^2)}\vert$ of time-like and space-like quark form factors: one could
 require (at least at large
$N_f$) that 
$4 C_F(H_{DY}^{new}(Q^2)-2H_{new}(Q^2))={d\over d\ln Q^2}\left[\ln\vert{{\cal F}(-Q^2)\over{\cal
F}(Q^2)}\vert^2\right]$. Thus, given a `natural' choice of $G_{DY}^{new}$, hence of $H_{DY}^{new}$, the
corresponding DIS functions
$H_{new}$ and $G_{new}$ would be fixed. An important issue is that of {\em universality} \cite{Lee} (here extended
between space-like and time-like processes): given an ansatz for a universal Sudakov effective coupling at large
$N_f$ (e.g. eq.(\ref{eq:A-simple})), one should check (work in progress) whether it remains universal at the
perturbative level at {\em finite}
$N_f$, using the {\em same} Sudakov distribution functions (e.g. eq.(\ref{eq:Gdot-SDG}) and (\ref{eq:G-DY-new}))
found at
$N_f=\infty$. Further insight 
could be afforded by a better understanding of OPE at large $N$ in the DIS case.
 Whatever the correct choice of the Sudakov distribution function
turns out to be,  it remains for  future phenomenological work  to determine the corresponding  form of the
(possibly non-perturbative) Sudakov  coupling at {\em finite} 
$N_f$ for each process, and test for eventual  deviations from universality. This step requires a parametrization
of the IR part of the  coupling, which can be viewed as an alternative to the shape function approach
\cite{Korchemsky}, but does not involve any explicit IR cut-off.

\noindent I thank M. Beneke,  Yu.L. Dokshitzer,  J-P. Lansberg, G. Marchesini, G.P. Salam and G. Sterman
for useful discussions. I am indebted to S. Friot for the result quoted in \cite{footnote5}. I also wish to thank
the referee for thought-stimulating comments.

\bibliography{apssamp}

\begin{thebibliography}{9}

\bibitem{DMW} Yu.L. Dokshitzer, G. Marchesini and B.R. Webber,  
{\em Nucl. Phys.} {\bf B469} 
(1996) 93, and references therein; Yu.L. Dokshitzer and B.R. Webber, Phys.Lett. {\bf B404} (1997) 321.

\bibitem{Gru-talk} G. Grunberg, hep-ph/0601140.
   
\bibitem{Vogt} S. Moch, J.A.M. Vermaseren and A. Vogt, Nucl.Phys. {\bf B726} (2005) 317.

\bibitem{GG} G. Grunberg, {\em Phys. Rev.} {\bf D29} 
(1984) 2315.


\bibitem{Gardi-Roberts} E. Gardi and R.G. Roberts, Nucl.Phys. {\bf B653} (2003) 227.



\bibitem{Beneke-Braun} M. Beneke and V.M. Braun, Nucl.Phys. {\bf B454} (1995) 253.




\bibitem{GS} G.P. Korchemsky and G. Sterman, Nucl.Phys. {\bf B437}
(1995) 415.







\bibitem{Gru-Borel} G. Grunberg, Phys.Lett. {\bf B304} (1993) 183.


\bibitem{footnote2} I am assuming   here, as well as in eq.(\ref{eq:scale-viol}), that the effective coupling
$A_{{\cal S}}(k^2)$ arising  directly  from Sudakov resummation is an {\em Euclidean} coupling (rather then
 the (integrated) time-like discontinuity of  an  Euclidean coupling \cite{DMW}, which might be
justified only in the single dressed gluon approximation), so that {\em all} power behaved terms (whether
analytic or not) in the low energy expansion of
$F(k^2/Q^2,N)$ do contribute to power corrections within the IR finite coupling framework.  
Identifying power corrections  only through    non-analytic terms   will also lead to a
conflict with the IR renormalon expectation if $B[A_{{\cal S}}](u)$ does not vanish at the renormalon position.


\bibitem{footnote3} The natural occurence of this combination of anomalous
dimensions has been also pointed out   by G.Sterman (talk given at the FRIF
Workshop on first-principles non-perturbative QCD of hadron jets (Paris, January 2006)).



\bibitem{BB} M. Beneke and V.M. Braun, Phys.Lett. {\bf B348} (1995) 513; P. Ball, M. Beneke and V.M. Braun, {\em
Nucl. Phys.} {\bf B452} (1995) 563.

\bibitem{Vogelsang} G. Sterman and W. Vogelsang, hep-ph/9910371; E. Laenen, G. Sterman and W. Vogelsang,
hep-ph/0010183; {\em Phys. Rev.} {\bf D63} (2001) 114018.

\bibitem{footnote} Consistently with the general  assumption \cite{footnote2} that $A_{{\cal S},DY}^{new}(k^2)$ is
an Euclidean coupling, I am discarding here (a similar remark applies below eq.(\ref{eq:A-simple})) the
alternative  interpretation that $G_{DY}^{new}(\epsilon_{DY})$ should be viewed as a `characteristic function'
\cite{DMW} where power corrections are related to the {\em discontinuities} of non-analytic terms at small
$\epsilon_{DY}$.


\bibitem{footnote4} An exception is the case $C_{new}(u)=B[A_{{\cal S}}^{new}](u)=1$, i.e. $\Gamma_{new}(u)={\pi
u\over \sin\pi u}$, which  corresponds to a one loop effective Sudakov  coupling.

\bibitem{footnote5} The alternative `minimal' ansatz $B[A_{{\cal S}}^{new}](u)=1$ fits less well within
the IR finite coupling approach, since it yields \cite{footnote4} an effective coupling with a Landau pole.
Moreover, the corresponding Sudakov distribution function does {\em not} exist in the DIS case, although it does 
in the Drell-Yan case, where one gets $G_{DY}^{new}(\epsilon_{DY})=J_0(2\epsilon_{DY})-1$ ($J_0$ is the Bessel
function).



\bibitem{Magnea} G. Sterman, Nucl.Phys. {\bf B281} (1987) 310; T.O. Eynck, E. Laenen and L. Magnea, {\em JHEP}
{\bf 0306} (2003) 057.  

\bibitem{Lee} C. Lee and G. Sterman, hep-ph/0603066, and references therein. 

\bibitem{Korchemsky} G.P. Korchemsky, hep-ph/9806537;   G.P. Korchemsky and G. Sterman, Nucl.Phys.   {\bf B555}
(1999) 335.

\end{thebibliography}

\end{document}